\documentstyle[aps,prl,preprint,floats,epsfig]{revtex}
\begin{document}

\setlength{\footnotesep}{11pt}

\title{\Large\bf Measurement of the
K$_L\rightarrow e^+e^- e^+e^-$ Decay Rate}
\vspace{8mm}
\author{\large \bf NA48 Collaboration\footnote{\small
Contact: Matthias Wittgen, Institut f\"ur Physik, Universit\"at Mainz, D-55099 Mainz.\\*[-3mm]
E-mail: Matthias.Wittgen@uni-mainz.de\\*[-5mm]}}
\date{\today}
\maketitle
\tighten
\vspace{8mm}

\begin{center}
\small{

A.~Lai, D.~Marras
\\
{\small \em Dipartimento di Fisica dell'Universit\`a e Sezione
    dell'INFN di Cagliari, I-09100 Cagliari, Italy.} \\[0.4cm]
A.~Bevan, R.S.~Dosanjh, T.J.~Gershon,
B.~Hay\footnote{\small Present address: EP Division, CERN, 1211 Gen\`eve 23, Switzerland.},
G.E.~Kalmus, C.~Lazzeroni, D.J.~Munday,
M.D.~Needham\footnote{\small Present address: NIKHEF, PO Box 41882, 1009 DB Amsterdam, The Netherlands.},
E.~Olaiya, M.A.~Parker, T.O.~White, S.A.~Wotton
\\
{\small \em Cavendish Laboratory, University of Cambridge, 
    Cambridge, CB3 0HE, U.K.\footnote{\small Funded by the U.K.
    Particle Physics and Astronomy Research Council.}} \\[0.4cm]
G.~Barr, G.~Bocquet, A.~Ceccucci, T.~Cuhadar, D.~Cundy, G.~D'Agostini, P.~Debu, N.~Doble, 
V.~ Falaleev, L.~Gatignon, A.~Gonidec, B.~Gorini, G.~Govi, P.~Grafstr\"om,
W.~Kubischta, A.~Lacourt,
M.~Lenti\footnote{\small On leave from Sezione dell'INFN di Firenze, I-50125 Firenze, Italy.},
A.~Norton, S.~Palestini, B.~Panzer-Steindel,
G.~Tatishvili\footnote{\small On leave from Joint Institute for Nuclear Research, Dubna, 141980, Russian Federation},
H.~Taureg, M.~Velasco, H.~Wahl
\\
{\small \em CERN, CH-1211 Gen\`eve 23, Switzerland.} \\[0.4cm]
C.~Cheshkov,
A.~Gaponenko, P.~Hristov,
V.~Kekelidze, D.~Madigojine,
N.~Molokanova,
Yu.~Potrebenikov, A.~Tkatchev, A.~Zinchenko
\\ 
{\small \em Joint Institute for Nuclear Research, Dubna, Russian
    Federation.}  \\[0.4cm]
%
%
I.~Knowles, V.~Martin, R.~Sacco, A.~Walker
\\
{\small \em Department of Physics and Astronomy, University of
    Edinburgh, JCMB King's Buildings, Mayfield Road, Edinburgh,
    EH9 3JZ, U.K.} \\[0.4cm]
%
%
M.~Contalbrigo, P.~Dalpiaz, J.~Duclos,
P.L.~Frabetti, A.~Gianoli, M.~Martini, F.~Petrucci, M.~Savri\'e
\\
{\small \em Dipartimento di Fisica dell'Universit\`a e Sezione
    dell'INFN di Ferrara, I-44100 Ferrara, Italy.} \\[0.4cm]
%
%
A.~Bizzeti\footnote{\small Dipartimento di Fisica
                       dell'Universita' di Modena e Reggio Emilia, via G. Campi 213/A I-41100 Modena, Italy},
M.~Calvetti, G.~Collazuol, G.~Graziani, E.~Iacopini
\\
{\small \em Dipartimento di Fisica dell'Universit\`a e Sezione
    dell'INFN di Firenze, I-50125 Firenze, Italy.} \\[0.4cm]
%
\newpage
H.G.~Becker, M.~Eppard, H.~Fox, K.~Holtz, A.~Kalter, K.~Kleinknecht, U.~Koch, L.~K\"opke,
P.Lopes da Silva, P.~Marouelli, I.~Pellmann, A.~Peters, B.~Renk, 
S.A.~Schmidt, V.Sch\"onharting, Y.~Schu\'e, R.~Wanke, A.~Winhart, M.~Wittgen
\\
{\small \em Institut f\"ur Physik, Universit\"at Mainz, D-55099
    Mainz, Germany\footnote{\small Funded by the German Federal Minister for
    Research and Technology (BMBF) under contract 7MZ18P(4)-TP2.}.} \\[0.4cm]
J.C.~Chollet, L.~Fayard, L.~Iconomidou-Fayard, J.~Ocariz, G.~Unal, I.~Wingerter-Seez
\\
{\small \em Laboratoire de l'Acc\'el\'eratur Lin\'eaire, IN2P3-CNRS,
Universit\'e de Paris-Sud, 91406 Orsay,
France\footnote{\small Funded by Institut National de Physique des Particules
et de Physique Nucl\'eaire (IN2p3), France}.} \\[0.4cm]
G.~Anzivino, P.~Cenci, E.~Imbergamo, P.~Lubrano, A.~Mestvirishvili, A.~Nappi,
M.~Pepe, M.~Piccini
\\
{\small \em Dipartimento di Fisica dell'Universit\`a e Sezione
    dell'INFN di Perugia, I-06100 Perugia, Italy.} \\[0.4cm]
%
%
P.~Calafiura, C.~Cerri, M.~Cirilli,
F.~Costantini, R.~Fantechi, S.~Giudici, I.~Mannelli,  
G.~Pierazzini, M.~Sozzi
\\
{\small \em Dipartimento di Fisica, Scuola Normale Superiore e Sezione
INFN di Pisa, I-56100 Pisa, Italy.} \\[0.4cm]
%
%
J.B.~Cheze, J.~Cogan, M.~De Beer, A.~Formica, R.~Granier de Cassagnac,
E.~Mazzucato, B.~Peyaud, R.~Turlay, B.~Vallage
\\
{\small \em DSM/DAPNIA - CEA Saclay, F-91191 Gif-sur-Yvette, France.} \\[0.4cm]
M.~Holder, A.~Maier, M.~Ziolkowski \\
{\small \em Fachbereich Physik, Universit\"at Siegen, D-57068 
Siegen, Germany\footnote{\small Funded by the German Federal Minister for
Research and Technology (BMBF) under contract 056SI74.}.} \\[0.4cm]
R.~Arcidiacono, C.~Biino, N.~Cartiglia, R.~Guida, F.~Marchetto, 
E.~Menichetti, N.~Pastrone \\
{\small \em Dipartimento di Fisica Sperimentale dell'Universit\`a e
    Sezione dell'INFN di Torino, \\ I-10125 Torino, Italy.} \\[0.4cm]
J.~Nassalski, E.~Rondio, M.~Szleper, W.~Wislicki, S.~Wronka
\\
{\small \em Soltan Institute for Nuclear Studies, Laboratory for High
    Energy Physics, \\ PL-00-681 Warsaw, Poland\footnote{\small 
   Supported by the Committee for Scientific Research grant 2P03B07615
    and using computing resources of the Interdisciplinary Center for
    Mathematical and 
    Computational Modelling of the University of Warsaw.}.} \\[0.4cm]
H.~Dibon, G.~Fischer, M.~Jeitler, M.~Markytan, I.~Mikulec, G.~Neuhofer,
M.~Pernicka, A.~Taurok, L.~Widhalm
\\
{\small \em \"Osterreichische Akademie der Wissenschaften, Institut
    f\"ur Hochenergiephysik, \\ A-1050 Wien, Austria\footnote{\small Funded by
the Austrian Ministery for Traffic and Research under the contract
GZ 616.360/2-IV GZ 616.363/2-VIII, and by the Fonds f\"ur Wissenschaft und Forschung
FWF Nr.~P08929-PHY}.}
}
\end{center}

\newpage

\begin{abstract}
The decay rate of the neutral long-lived  K meson into 
the $e^+ e^-e^+e^-$ final state has been measured with the 
NA48 detector at the CERN SPS. Using data collected in 1999,
a total of 132 events has been observed with negligible
background. The total number of kaons  was determined to be $5.1\times 10^{10}$.
This observation corresponds to a preliminary branching ratio of
$\Gamma($K$_L\to e^+e^-e^+e^-)/\Gamma($K$_L\to \;$all$)= (3.67 \pm 0.32_{stat} \pm 0.23_{sys} \pm 0.08_{norm})\times 10^{-8}$,
where the first error is statistical, the 
second systematic and the third error is due to the uncertainty in the 
normalization.
\end{abstract}

\section{Introduction}

The K$_L \to e^+ e^- e^+ e^-$ decay is expected to proceed
mainly via the intermediate state 
K$_L\to \gamma^*\gamma^*$~\cite{bib:quigg,bib:gg,bib:miyazaki} 
and thus depends on the structure of the K$_L\rightarrow\gamma^*\gamma^*$ vertex. 
Phenomenological models include vector meson dominance of the photon 
propagator~\cite{bib:sak}, QCD inspired models~\cite{bib:shif}, intermediate
pseudoscalar and vector mesons~\cite{bib:bms} and models based on 
chiral perturbation theory~\cite{bib:chir}. 
The probability for both virtual photons to convert into  
$e^+e^-$ pairs is calculated to be in the range
$(5.89 - 6.50)\times 10^{-5}$~\cite{bib:miyazaki,bib:longzhe}.
The chiral model prediction of~\cite{bib:longzhe} corresponds to 
$\Gamma$(K$_L\rightarrow e^+e^-e^+e^-)/\Gamma($K$_L\rightarrow \;$all$) = 3.85 \times 10^{-8}$,
including the effect of a form factor, which increases the width by 4\%. The interference
term due to the identity of particles has been calculated to change the 
branching ratio by 0.5\%.  

The decay was first observed by the CERN NA31  
experiment~\cite{bib:eeeena31a} based on 2 observed events and has 
been confirmed by 
later measurements~\cite{bib:eeee}.


Here we report the preliminary result obtained from the  1999 data
taking of the CERN experiment NA48. 

\section{Experimental Setup and Data Taking}

This measurement was carried out as part of the CERN experiment NA48 at the 
CERN SPS, which 
has previously reported measurements of the related decays 
K$_L\rightarrow e^+e^-\gamma$~\cite{bib:eeg} and
K$_L\rightarrow \mu^+\mu^-\gamma$~\cite{bib:mmg}.
A detailed and comprehensive description of the detector is in 
preparation~\cite{bib:na48det}.

\begin{figure}[htb]
\begin{center}
\epsfig{file=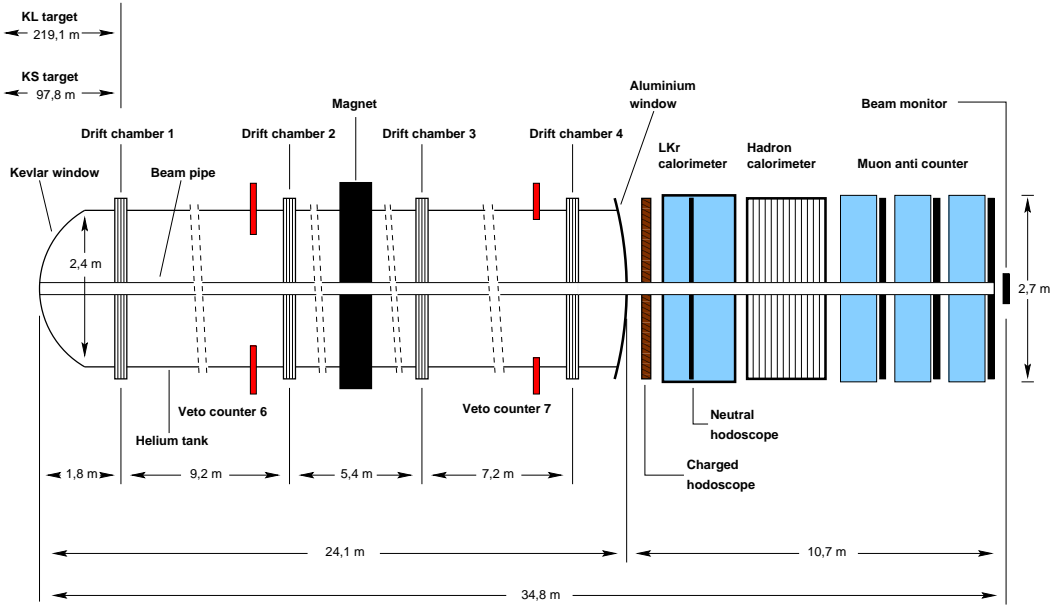,width=0.85\textwidth}
\vspace{4mm}
\caption{\sl Side-view of the NA48 detector.}
\vspace{-4mm}
\label{fig:na48}
\end{center}
\end{figure}

The NA48 experiment is designed specifically to measure the direct 
CP violation para\-meter $\Re{(\epsilon'/\epsilon)}$ using simultaneous beams 
of K$_L$ and K$_S$. To produce the K$_L$ beam, 450~GeV/c protons are
extracted from the accelerator during 2.4~s every 14.4~s and 
$1.1 \times 10^{12}$
of these are delivered to a beryllium target. Using dipole magnets to sweep
away charged particles and collimators to define a narrow beam, 
a  neutral beam of $2 \times 10^7$ K$_L$ per burst
and divergence $\pm 0.15$ mrad enters the decay region. The fiducial 
volume begins 126 m downstream from the target and 
is contained in an evacuated cylindrical steel vessel 89~m long and 2.4~m in
maximum diameter. 
The vessel is terminated at the downstream end by a Kevlon-fiber composite
window and followed immediately by the main NA48 detector. 
The sub-detectors which are used in the K$_L\rightarrow e^+e^- e^+e^-$ 
analysis are described below in order of succession (see Fig.~\ref{fig:na48}). 

A magnetic spectrometer consisting of a dipole magnet is preceded and followed 
by two sets of drift chambers. The drift chambers are each comprised of 
eight planes of sense wires, two horizontal, two vertical and two
along each of the 45$^\circ$ directions. Only 
the vertical and horizontal planes are instrumented in the third 
chamber. The volume between 
the chambers is filled with helium at atmospheric pressure. 
The momentum resolution is
$\Delta p/p = 0.65\%$ at 45~GeV/c.  

Two  segmented plastic scintillator hodoscope planes are placed after the 
helium tank and provide signals for the trigger.
 
A liquid krypton filled calorimeter (LKr) is used for measuring the energy, 
position and timing of electromagnetic showers. Spacial and timing resolutions
of better than 1.3 mm~and 300~ps, respectively, have been achieved
for energies above 20~GeV. The 
energy resolution is $\frac{\sigma (E)}{E} = \frac {0.035}{\sqrt E} \oplus 
\frac{0.110}{E} \oplus 0.006$, with $E$ measured in GeV. 

A hadron calorimeter composed of 48 steel plates, each 24~mm thick, interleaved
with scintillator is used in trigger formation and particle detection studies.

\section{Trigger and Data Processing}

Candidate events were selected by a two-stage trigger. 
At the first level, a trigger requiring adjacent hits in the 
hodoscope is put in coincidence with a total energy condition ($\ge 35$ GeV), 
defined by adding  the energy deposited in the hadronic calorimeter with 
that seen by the trigger in the LKr calorimeter. 
The second level  trigger  uses information from the drift chambers to 
reconstruct tracks and invariant masses.
For the 4-track part of the trigger, the number of clustered hits in each
of the first, second, and fourth drift chamber
had to be between 3 and 7. 
All possible 2-track vertices were calculated online. At least two vertices
within 6 m of each other in the axial direction had to be found.


For the determination of the 4-track trigger efficiency, downscaled
events that passed the first level were used. Alternatively, events 
triggered with the neutral trigger - based on the data of the 
LKr calorimeter - were selected. The neutral trigger 
applied the following cuts to the events online: $\le$ 5 peaks in 
each projection, total energy $>$ 50 GeV, first moment of cluster energies
(`center of gravity') $< 15 $ cm, and lifetime uncorrected for 
deflection in the magnet $< 4.5\cdot \tau_{K_S}$.

During the experimental runs, roughly 100 Terabytes of raw data with 
typically 20 kbytes per event were recorded.
Reconstructed output was stored in a compressed data format, 45 times
smaller. In addition, several streams of data were formed for events 
accepted by about 40 filter modules for the analysis of neutral and 
charged two-pion decays, rare decays, and events for the detector calibration. 
 
\section{Data Analysis}

The data sample which yields  K$_L\rightarrow e^+e^-e^+e^-$ also has been 
used to select K$_L\rightarrow \pi^+\pi^-\pi^0_{Dalitz}$ 
and K$_L\rightarrow \pi^0\pi^0_{Dalitz}\pi^0_{Dalitz}$ 
normalization events, with $\pi^0_{Dalitz}\rightarrow e^+e^-\gamma$.
The vertex was 
reconstructed from the 4 tracks by requiring that the sum of the 
squared transverse distances from the transverse vertex position,
weighted by the inverse track momentum,  be minimal. 

Events were preselected by 
requiring two positive and two negative tracks with 
distance of closest approach to the vertex $< 15$~cm.


All clusters in the LKr were required to be in a fiducial area given 
by an octogon about 5 cm smaller than the outside perimeter of the 
calorimeter and an inner radius of 15 cm.  
 The distance to dead calorimeter cells
had to exceed 2 cm to ensure negligible energy loss. 
The separation between cluster centers was required to be $>$ 3 cm.  
The energy of each cluster was required to exceed 2 GeV, well above the 
detector noise of 0.11 GeV per cluster. 

Electron candidates were identified by requiring that cluster centers
in the LKr be within 1.5 cm of the calculated shower maxima based upon the 
extrapolation of each track. The efficiency of this procedure was 
measured to be $(99.7\pm 0.1)\%$~\cite{bib:eeg}. 
To reject pion showers, the ratio of cluster energy to track 
momentum $E/p$ was required to lie between 
0.9 and 1.2. The efficiency of this selection was determined to be 
$\ge 95\%$~\cite{bib:eeg}.
Those track-associated clusters with $0 < E/p < 0.8$ 
were classified as pions.
From a study of K$_S\rightarrow \pi^+\pi^-$ decays, the probability of pions 
to be wrongly classified as electrons is estimated to be 0.9\%;  the pion
classification is passed by 97.5\% of all pions.

The fiducial volume was defined by the axial postion  7.50~m~$< z_{vertex} < 90$~m downstream of the K$_S$ target. 
Within this volume, 4-track vertices were determined
with a typical axial resolution of  0.5 m, as estimated by the MC. 
The K$_L$ energy had to be in the range 50~GeV$-$200~GeV.

\subsection{Background Rejection and 
Selection of K$_L\rightarrow e^+e^-e^+e^-$ Candidates}

\begin{figure}[htb]
\begin{center}
\epsfig{file=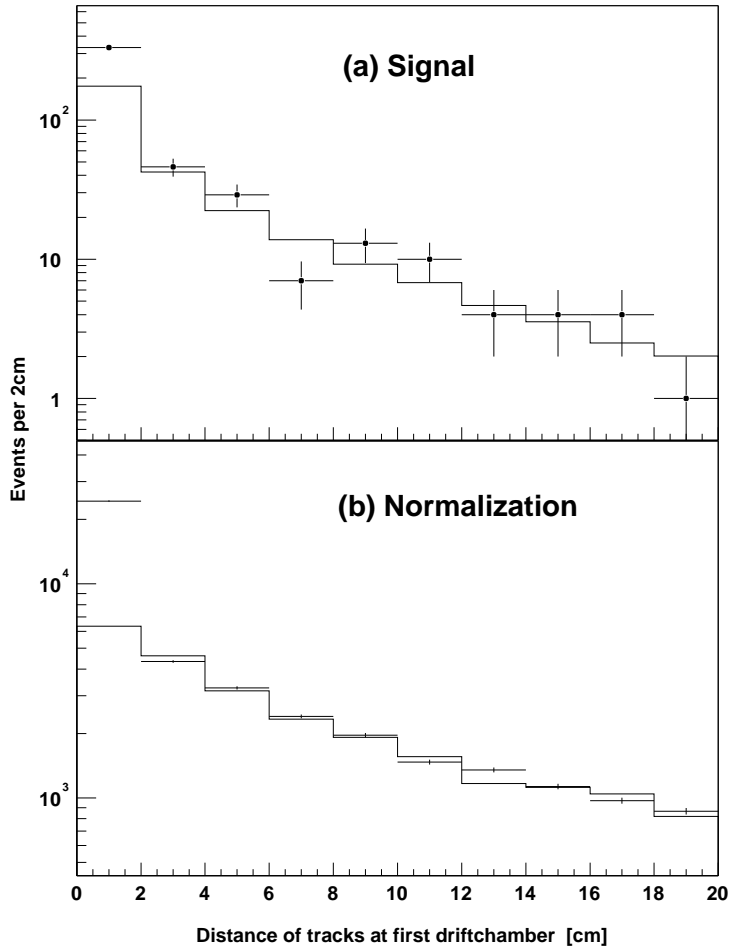,width=0.65\textwidth}
\vspace{4mm}
\caption{\sl Distance $d$ of each pair of oppositely charged tracks
at the position 
of the first drift chamber for (a) K$_L\rightarrow e^+e^-e^+e^-$ and 
(b) K$_L\rightarrow \pi^+\pi^-\pi^0_{Dalitz}$.
Dots with error bars show the data, 
the histogram is the Monte Carlo prediction normalized to the number 
of observed signal events with $d > 2$ cm.}
\vspace{-4mm}
\label{fig:distdch}
\end{center}
\end{figure}

Candidate events for the decay K$_L\rightarrow e^+e^-e^+e^-$  
with all tracks identified as electrons were selected. 
In principle, the following four classes of 
background sources are
relevant:
\begin{itemize}
\item 
Events with two decays 
K$_L\rightarrow \pi e \nu$  occuring at the 
same time and for which the pions were misidentified as electrons.
Being due to two coincident kaon decays the invariant mass of the system 
can be around and above the nominal K$_L$ mass. 
These events are largely rejected by requiring a good vertex quality:
the closest distance of approach of each track w.r.t. the reconstructed vertex
had to be  $<$ 5 cm. Events with separate vertices do not pass
the level 2 trigger (see above).  
 
Finally it was required that the measured times for  clusters associated 
to the electrons 
had to be consistent within 3 ns with each other. 
A study of sidebands
in this time distribution shows that the background from this source 
is negligible.

\item
Events K$_L\rightarrow \pi^0\pi^0\pi^0$, where the $\pi^0$'s undergo
single or double Dalitz decays or photons convert in the 
material of the detector, so that 2 positive and 2 negative 
electrons are detected. Due to the missing photons, the invariant 
mass of the $e^+e^-e^+e^-$ system is below the nominal K$_L$ mass.   
The loose requirement that the 
square of the reconstructed transverse momentum $p_T^2$ of the reconstructed 
kaon with respect to the line joining the decay vertex and the K$_L$ target
must be less than 0.0005 (GeV/c)$^2$, restricts this background to 
less than 2.2\% .
The position of the cut is indicated in Fig.~\ref{fig:pt2}.
The Monte Carlo simulation indicates  
that  1.5\% of the signal events are lost by  the $p_T^2$  cut. 

\begin{figure}[htb]
\begin{center}
\epsfig{file=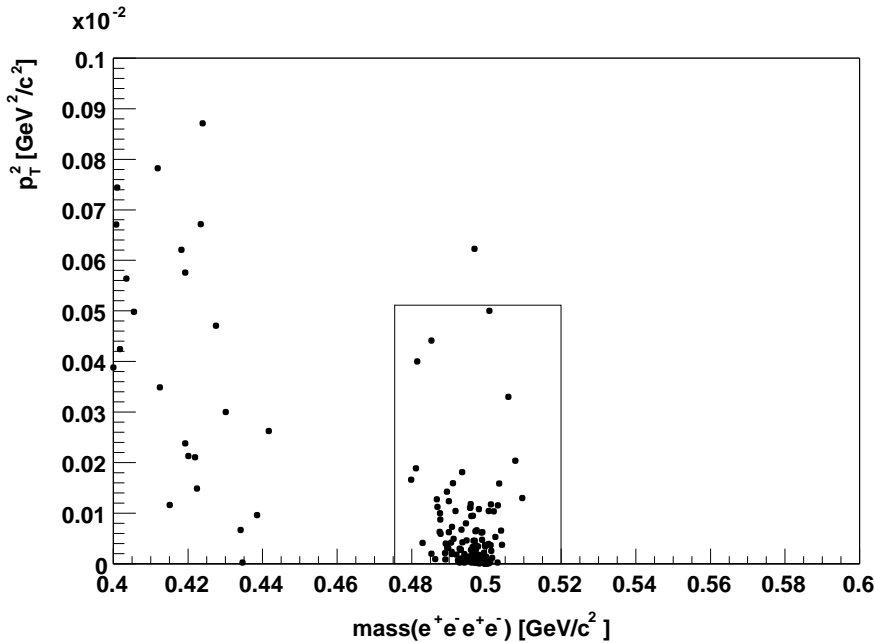,width=0.7\textwidth}
\end{center}
\caption{\sl Correlation of $e^+e^-e^+e^-$ invariant mass with 
the squared transverse momentum $p_T^2$ of 
the reconstructed kaon.}
\label{fig:pt2}
\end{figure}

\item
Events K$_L\rightarrow \gamma \gamma$ and 
K$_L\rightarrow e^+e^-\gamma$, with conversion of the photons in the 
material upstream of the spectrometer also yield invariant masses around 
the nominal K$_L$ mass. 
Each pair of oppositely charged tracks was therefore 
required to be separated 
by $\ge 2$~cm in the first drift chamber as indicated in  
Fig.~\ref{fig:distdch}(a).   
Note that the conversion probability in the material of the NA48 detector
is of similar magnitude as that for internal photon conversion to a 
$e^+e^-$ pair.
As the angular opening of oppositely 
charged tracks peaks  mostly at small angles, 60\% of the signal 
events are lost by this cut; according to the MC, 
there is no remaining background with converted $\gamma$'s.
\item 
Events K$_L\rightarrow \pi^+\pi^-e^+e^-$~\cite{bib:Kppee},
with the pions misidentified as electrons.
Due to the low misidentification probability of 0.9\% 
this background is found to be negligible. 
\end{itemize}

\begin{figure}[htb]
\begin{center}
\epsfig{file=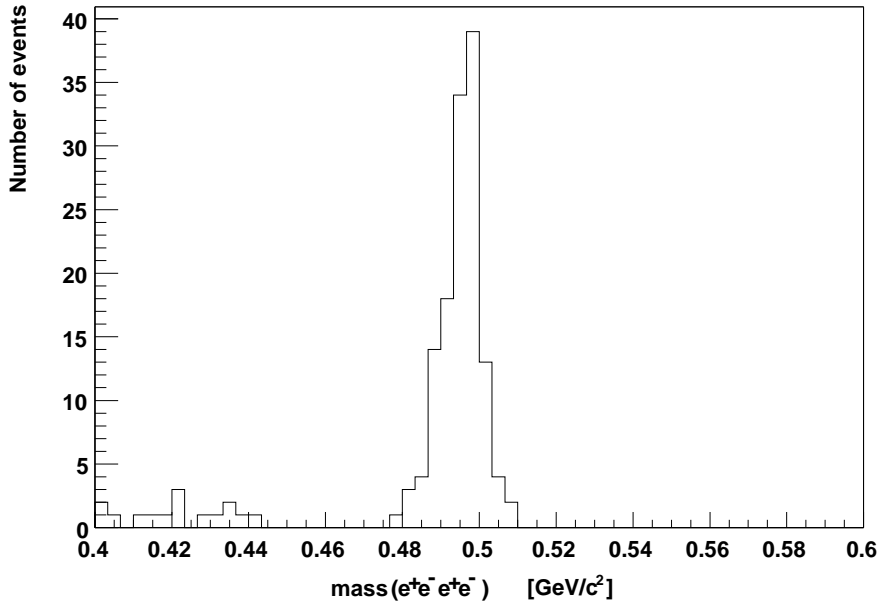,width=0.7\textwidth}
\vspace{4mm}
\caption{\sl Invariant mass of the $e^+e^-e^+e^-$ system.}
\vspace{-4mm}
\label{fig:massKL}
\end{center}
\end{figure}

The invariant $e^+e^-e^+e^-$ mass plot resulting from this selection 
is shown in 
Fig.~\ref{fig:massKL}. Note the slightly asymmetric shape of the K$_L$ mass 
peak, which is due to photons radiated off the electrons in the 
final state. 

Finally, a mass window of 
0.475 GeV/$c^2 <  m(e^+e^-e^+e^-)  <$ 0.520 GeV/$c^2$  was set to define the 
final sample. 
In total, 132 candidate events were selected in this way. 
From the number of events observed below the K$_L$ mass peak
and their distribution
we estimate that the background contribution to the signal region is
negligible.


\subsection{Normalization}

The four-track decay  K$_L\rightarrow \pi^+\pi^-\pi^0_{Dalitz}$, with 
$\pi^0_{Dalitz}\rightarrow e^+e^-\gamma$, was used for norma\-li\-zation. 
Only events that  passed the pretrigger discussed above, downscaled by 60, 
were selected.
 This sample was used to determine the efficiency of the 
4-track trigger to be $95.1\pm 0.2\%$.  
The 4-track trigger 
is required for signal and normalization modes.
 
Since both the signal and the normalization modes consist of 
4-track events, uncertainties due to tracking tend to cancel in the 
ratio of acceptances.  Selection criteria
similar to those used in the signal mode
were applied. In addition, at least one extra cluster in the calorimeter
not associated with a charged track was required. The invariant mass of 
the $e^- e^-\gamma$ system had to be in the range of 0.115 -- 0.155 
GeV/$c^2$.  
Monte Carlo studies showed that 66 \% of the reconstructed 
K$_L\rightarrow \pi^+\pi^-\pi_D^0$ candidate decays are from 
K$_L\rightarrow \pi^+\pi^-\pi^0$ with one of the external photons converting
in the material of the detector (see Fig.~\ref{fig:distdch}(b)).
All other backgrounds have been estimated
to be negligible.
Using these cuts, 17123  
K$_L\rightarrow \pi^0\pi^0\pi^0_D$ decays were selected.    

In a second analysis, K$_L\rightarrow \pi^0\pi^0_{Dalitz}\pi^0_{Dalitz}$ events 
were selected, yielding 5167 events for normalization.
While having a complicated topology of eight clusters,  
these events have the advantage that all decay products interact 
electromagnetically in the detector and that the radiative
corrections should be similar to those in the signal mode.
 
\subsection{Acceptance Determination and Kaon Flux}

For the simulation of the K$_L\rightarrow e^+e^-e^+e^-$ acceptance, 
the matrix element was taken from Ref.~\cite{bib:miyazaki}
neglecting the interference
of the two virtual photons. The distribution of the angle 
spanned by the decay planes of the 
two $e^+e^-$ pairs corresponds to a K$_L$ which is assumed to be entirely CP$= -1$. 
The PHOTOS~\cite{bib:PHOTOS} package has been used 
to simulate final state radiation both for the signal and normalization 
channels.

The acceptance for K$_L\rightarrow e^+e^-e^+e^-$ is calculated to be
7.8\% for events generated in the range 45 GeV $ < E_{K_L} < 215$ 
GeV and 5~m~$< z_{vertex} < 91$~m. 
The normalization to the K$_L$ signal has been measured from the number
of K$_L\rightarrow\pi^+\pi^-\pi^0_{Dalitz}$ decays in the same sample. Using 
the acceptance of 1.34\%, calculated by Monte Carlo simulation, 
a total number of  $5.1 \times 10^{10}$ K$_L$ decays is obtained. 
With the cuts described above, the inclusion of 
radiative corrections decreased the 
acceptance of signal  and normalization channel by 8.8\% and 3.4\%, 
respectively.

Consistent results for the total number of K$_L$  were obtained when the 
alternative normalization channel, K$_L\rightarrow \pi^0\pi^0_{Dalitz}
\pi^0_{Dalitz}$, was used instead.  


\section{Results and Discussion}
A study of the stability of the branching ratio determination was made
as a function of the cuts applied. A systematic error of 3.5\% was 
estimated, mainly being due to the variation in the minimal distance 
of clusters from the beam pipe and the cut on the vertex quality. 

A second contribution comes from the 4-track trigger inefficiency.
A Monte Carlo si\-mu\-la\-tion of the level 2 algorithm yields 
a 95.0\% efficiency for the normalization mode, in good agreement 
with the measured value. For the signal, the simulated efficiency 
is higher (99.8\%). We chose not to apply a correction to the branching ratio;
instead  we introduced a systematic error of $\pm$5\% for this 
preliminary result.  
 
Finally, the effect of overflows in the drift chambers has been considered. 
If events with an overflow condition in a window of 312 ns around 
the event time are removed, 20\% of the events are lost and the 
branching ratio stays constant within 1\%.  

Adding these sources of systematic error in quadrature, we obtain 
a total systematic error of $\pm$ 6.2\%.

From the numbers given above, a branching ratio of
\[
\frac{\Gamma({\rm K}_L\rightarrow e^+e^-e^+e^-)}{\Gamma({\rm K}_L\rightarrow {\rm all})} =
(3.67 \pm 0.32_{stat} \pm 0.23_{syst} 
\pm 0.08_{norm} )\times 10^{-8}
\]
is obtained, where the statistical and systematical uncertainties as well 
as the uncertainty in the branching ratio of the normalization channel
are given
separately.
 
This result is consistent with the theoretical expectation 
of~\cite{bib:longzhe} and the previous average value 
of ($4.1\pm 0.8)\times 10^{-8}$~\cite{bib:PDG},
with 5 times the statistics of the single best previous experiment. 

. 
   
%
%

\section*{Acknowledgements}
It is a pleasure to thank the technical staff of the participating laboratories,
universities and affiliated computing centers for their efforts in the 
construction of the NA48 apparatus, in operation of the experiment, and in 
the processing of the data.

\end{document}